\documentclass{aastex}
\usepackage{spr-astr-addons}

\begin{document}

\title{WD + He star systems as the progenitors of Type Ia supernovae and their surviving companion stars
}

\shorttitle{Progenitors of SNe Ia and their companions}
\shortauthors{Wang \& Han}

\author{B. Wang\altaffilmark{1,2} }
\and \author{Z. Han\altaffilmark{1}} \email{wangbo@ynao.ac.cn}

\altaffiltext{1}{National Astronomical Observatories/Yunnan
Observatory, Chinese Academy of Sciences,
           Kunming 650011, China
           \\E-mail: wangbo@ynao.ac.cn}
\altaffiltext{3}{Graduate University of the Chinese Academy of
Sciences, Beijing 100049, China}

\begin{abstract}
Employing Eggleton's stellar evolution code with an optically thick
wind assumption, we have systematically studied the WD + He star
channel of Type Ia supernovae (SNe Ia), in which a carbon-oxygen WD
accretes material from a He main-sequence star or a He subgiant to
increase its mass to the Chandrasekhar mass. We mapped out the
parameter spaces for producing SNe Ia. According to a detailed
binary population synthesis approach, we find that the Galactic SN
Ia birthrate from this channel is $\sim$$0.3\times 10^{-3}\ {\rm
yr}^{-1}$, and that this channel can produce SNe Ia with short delay
times ($\sim$45$-$140\,Myr). We also find that the surviving
companion stars in this channel have a high spatial velocity
($>$400\,km/s) after SN explosion, which could be an alternative
origin for hypervelocity stars (HVSs), especially for HVSs such as
US 708.

\end{abstract}


\section{Introduction}
Type Ia supernovae (SNe Ia) play an important role in the study of
cosmic evolution, especially in cosmology. They have been applied
successfully in determining cosmological parameters (e.g. $\Omega$
and $\Lambda$; Riess et al. 1998; Perlmutter et al. 1999). It is
generally believed that SNe Ia are thermonuclear explosions of
carbon-oxygen white dwarfs (CO WDs) in binaries. However, there is
still no agreement on the nature of their progenitors (Hillebrandt
and Niemeyer 2000; Podsiadlowski et al. 2008; Wang et al. 2008).

Over the past few decades, two families of SN Ia progenitor models
have been proposed, i.e. the double-degenerate (DD) and
single-degenerate (SD) models. Of these two models, the SD model is
widely accepted at present. It is suggested that the DD model, which
involves the merger of two CO WDs (Iben \& Tutukov 1984; Webbink
1984; Han 1998), likely leads to an accretion-induced collapse
rather than to an SN Ia (Nomoto and Iben 1985). For the SD model,
the companion is probably a main-sequence (MS) star or a slightly
evolved subgiant star (WD + MS channel), or a red-giant star (WD +
RG channel) (Hachisu et al. 1996; Li and van den Heuvel 1997; Langer
et al. 2000; Han and Podsiadlowski 2004, 2006; Chen and Li 2007,
2009; L\"{u} et al. 2009; Meng et al. 2009; Wang, Li and Han 2009;
Meng and Yang 2009a). Meanwhile, a CO WD may also accrete material
from a He star to increase its mass to the Chandrasekhar (Ch) mass,
which is known as the WD + He star channel. Yoon and Langer (2003)
followed the evolution of a CO WD + He star system, in which the WD
can increase its mass to the Ch mass by accreting material from the
He star. It is believed that WD + He star systems are generally
originated from intermediate mass binaries, which may explain SNe Ia
with short delay times implied by recent observations (Mannucci et
al. 2006, Aubourg et al. 2008).

The purpose of this paper is to study SN Ia birthrates and delay
times of the WD + He star channel and to explore the properties of
the surviving companion stars after SN explosion. In Section 2, we
describe the numerical code for the binary evolution calculations
and the binary evolutionary results. We describe the binary
population synthesis (BPS) method and results in Section 3. Finally,
a discussion is given in Section 4.

\section{Binary evolution calculations}

\begin{figure}[tb]
\includegraphics[width=5.5cm,angle=270]{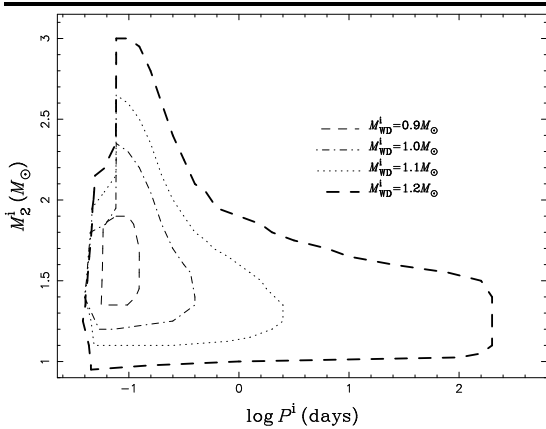}
 \caption{Regions in orbital
period--secondary mass plane for WD binaries that produce SNe Ia for
initial WD masses of $0.9, 1.0, 1.1$ and $1.2$\,$M_{\odot}$. The
lowest WD mass for producing SNe Ia in this channel is
0.865\,$M_{\odot}$.}
\end{figure}

We use Eggleton's stellar evolution code (Eggleton 1971, 1972, 1973)
to calculate the binary evolutions of WD + He star systems. The code
has been updated with the latest input physics over the past three
decades (Han et al. 1994; Pols et al. 1995, 1998). Roche lobe
overflow (RLOF) is treated within the code described by Han et al.
(2000). We set the ratio of mixing length to local pressure scale
height, $\alpha=l/H_{\rm p}$, to be 2.0. The opacity tables are
compiled by Chen and Tout (2007). In our calculations, He star
models are composed by He abundance $Y=0.98$ and metallicity
$Z=0.02$.

Instead of solving stellar structure equations of a WD, we use an
optically thick wind model (Hachisu et al. 1996) and adopt the
prescription of Kato and Hachisu (2004) for the mass accumulation
efficiency of He-shell flashes onto the WD. We have calculated about
2600 WD + He star systems, and obtain a large, dense model grid (for
details see Wang et al. 2009a). In Fig. 1, we show the contours for
producing SNe Ia. If the parameters of a WD binary at the onset of
the RLOF are located in the contours, an SN Ia is then assumed to be
produced. Thus, these contours can be expediently used in BPS
studies.

\section{Binary population synthesis}

To obtain SN Ia birthrates and delay times of the WD + He star
channel, we performed a series of Monte Carlo simulations in the BPS
study. In the simulation, by using the Hurley's rapid binary
evolution code (Hurley et al. 2000, 2002), we followed the evolution
of $4\times10^{\rm 7}$ sample binaries from the star formation to
the formation of the WD + He star systems according to the SN Ia
production regions (Fig. 1) and three evolutionary channels (i.e.
the He star channel, the EAGB channel, and the TPAGB channel; for
details see Wang et al. 2009b). Here, we adopt the standard energy
equations to calculate the output of the common envelop (CE) phase
(e.g. Wang et al. 2009b).

In the BPS study, the primordial binary samples are generated in the
Monte Carlo way and a circular orbit is assumed for all binaries. We
adopt the following input for the simulation (e.g. Han et al. 2002,
2003, 2007; Wang et al. 2009b). (1) The initial mass function (IMF)
of Miller and Scalo (1979) is adopted. (2) The mass-ratio
distribution is taken to be constant. (3) The distribution of
separations is taken to be constant in $\log a$ for wide binaries,
where $a$ is the orbital separation. (4) We simply assume a constant
star formation rate (SFR) over the past 15\,Gyr or, alternatively,
as a delta function, i.e. a single starburst.

\begin{figure}[tb]
\includegraphics[width=5.5cm,angle=270]{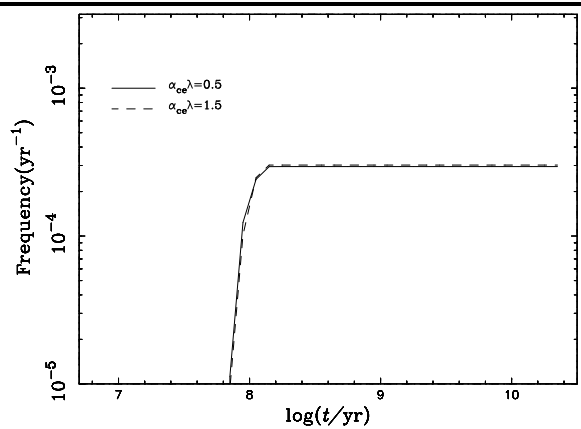}
 \caption{The evolution of Galactic SN Ia birthrate for a constant Pop I
SFR ($5\,M_{\odot}$\,yr$^{-1}$). The solid and dashed curves show
the results of different CE ejection parameters with $\alpha_{\rm
ce}\lambda=0.5$ (solid) and $\alpha_{\rm ce}\lambda=1.5$ (dashed),
respectively. }
\end{figure}

\begin{figure}[tb]
\includegraphics[width=5.5cm,angle=270]{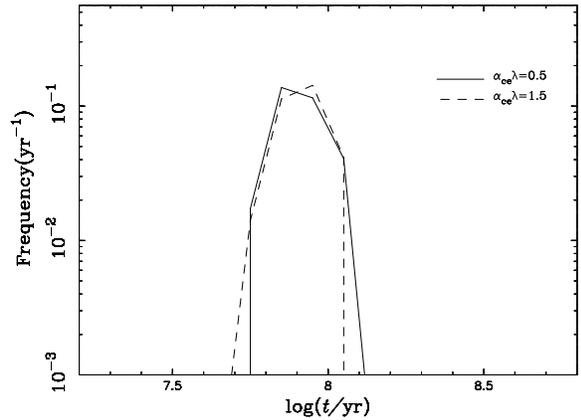}
 \caption{Similar to Fig. 2, but for a single starburst with a total mass of $10^{\rm
11}\,M_{\odot}$.}
\end{figure}

In Fig. 2, we show Galactic birthrates of SNe Ia for the WD + He
star channel by adopting $Z=0.02$ and ${\rm SFR}=5\,M_{\rm
\odot}{\rm yr}^{-1}$. The simulations give Galactic SN Ia birthrate
of $\sim$$0.3\times 10^{-3}\ {\rm yr}^{-1}$. Figure 3 displays the
evolution of SN Ia birthrates for a single starburst with a total
mass of $10^{11}\,M_{\odot}$. In the figure, we see that SN Ia
explosions occur between $\sim$45\,Myr and $\sim$ 140\,Myr after the
starburst, which may explain SNe Ia with short delay times.

\begin{figure}[tb]
\includegraphics[width=8.4cm,angle=0]{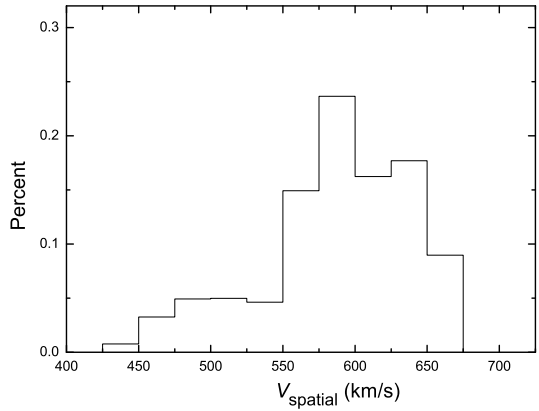}
 \caption{The distribution of the spatial velocity with
 $\alpha_{\rm ce}\lambda=0.5$.}
\end{figure}

The companion star in the SD model would survive in the SN explosion
and potentially be identifiable (Podsiadlowski 2003; Han 2008; Meng
and Yang 2009b). We obtained the distributions of many properties of
the companion stars of this channel at the moment of SN explosion
(e.g. Wang and Han 2009). We can give the spatial velocity of the
surviving companion stars, based on the formula $V_{\rm 2}^{\rm
SN}=\sqrt{V_{\rm kick}^{2}+V_{\rm orb}^{2}}$, where $V_{\rm kick}$
and $V_{\rm orb}$ are the kick velocity and the orbital velocity of
the companion star at the moment of SN explosion, respectively. The
kick velocity depends on the ratio of separation to the radius of
companions at the moment of SN explosion, $A/R_{\rm 2}^{\rm SN}$,
and the leading head velocity of SN ejecta (Meng et al. 2007). Here,
the leading head velocity is assumed to be 13500\,km/s, which is
from the SN ejecta kinetic energy $1.5\times10^{51}$\,erg
corresponding to the upper limit of the kinetic energy of normal SNe
Ia (Gamezo et al. 2003). In Fig. 4, we show the current epoch
distribution of the spatial velocity for the surviving companions
from this channel. We see that the surviving companion stars have
high spatial velocities ($>$400\,km/s), which almost exceed the
gravitational pull of the Galaxy nearby the sun. Thus, the surviving
companion stars from the WD + He star channel could be an
alternative origin for hypervelocity stars (HVSs), which are stars
with a velocity so great that they are able to escape the
gravitational pull of the Galaxy.

\section{Discussion}

The simulations give Galactic SN Ia birthrate of $\sim$$0.3\times
10^{-3}\ {\rm yr}^{-1}$, which is lower than that inferred
observationally (i.e. $3 - 4\times 10^{-3}\ {\rm yr}^{-1}$;
Cappellaro and Turatto 1997). This implies that the WD + He star
channel is only a subclass of SN Ia production, and there may be
some other channels or mechanisms also contributing to SNe Ia, e.g.
WD + MS channel, WD + RG channel or double-degenerate channel (see
Wang, Li and Han 2009; Meng and Yang 2009).

In this paper, we assume that all stars are in binaries and about
50\% of stellar systems have orbital periods less than 100\,yr. If
we adopt 46.3\% of stellar systems have orbital periods below
100\,yr by adjusting the parameter $a_{\rm 1}$ in equation (7) of
Wang et al. (2009b), the SN Ia birthrate from this channel will
decrease to be $\sim 0.28\times10^{-3}\,{\rm yr}^{-1}$.

Hachisu et al. (2008) investigated new evolutionary models for SN Ia
progenitors, introducing the mass-stripping effect on a MS or
slightly evolved companion star by winds from a mass-accreting WD.
The model can also provide a possible ways of producing young SNe
Ia, but the model depends on the efficiency of the mass-stripping
effect. We also find that the model produces very few young SNe Ia
according to a detailed BPS approach. Thus, we consider the WD + He
star channel as a main contribution to the formation of young SNe
Ia.


US 708 is an extremely He-rich sdO star in the Galaxy halo, with a
heliocentric radial velocity of +$708\pm15$\,${\rm km/s}$ (Hirsch et
al. 2005). We note that the local velocity relative to the Galatic
center may lead to a higher observation velocity for the surviving
companion stars, but this may also lead to a lower observation
velocity. Considering the local velocity near the sun
($\sim$220\,km/s), we find that $\sim$30\% of the surviving
companion stars may be observed to have velocity $V>700\,{\rm km/s}$
for a given SN ejecta velocity 13500\,km/s. In addition, the
asymmetric explosion of SNe Ia may also enhance the velocity of the
surviving companions. Thus, a surviving companion star in the WD +
He star channel may have a high velocity like US 708 (see also
Justham et al. 2008).

For a single starburst, most of the SN explosions occur between
$\sim$45\,Myr and $\sim$ 140\,Myr after the starburst, i.e. SNe Ia
from the WD + He star channel will be absent in the old galaxies. In
future investigations, we will employ the Large sky Area
Multi-Object fiber Spectral Telescope (LAMOST) to search the HVSs
originating from the surviving companions of SNe Ia.

\acknowledgments We thank the referee for valuable comments that
helped us to improve the paper. B.W. thanks Dr. S. Justham for
stimulating discussions. This work is supported by the National
Natural Science Foundation of China (Grant No. 10821061) and the
National Basic Research Program of China (Grant No. 2007CB815406).


\end{document}